\documentclass[superscriptaddress,floatfix,twocolumn,letterpaper]{revtex4}
\usepackage{graphicx}
\newcommand{\beq}{\begin{eqnarray}}
\newcommand{\eeq}{\end{eqnarray}}

\newcommand{\la}{\langle}
\newcommand{\ra}{\rangle}

\begin{document}
%
\title{Lyapunov instability of rigid diatomic molecules in three dimensions - a simpler method}

\author{Seungho Choe}
\affiliation{Department of Mechanical Engineering, University of
Michigan, 2350 Hayward Street, Ann Arbor, MI 48109, USA}

\author{Eok-Kyun Lee}
\affiliation{Department of
Chemistry, Korea Advanced Institute of Science and Technology,
373-1 Kusung-dong, Yusung-ku, Taejon, 305-701, Korea}


\begin{abstract}
We present a new method to calculate Lyapunov exponents of rigid
diatomic molecules in three dimensions ($12N$ dimensional phase
space). The spectra of Lyapunov exponents are obtained for 32 rigid
diatomic molecules interacting through the
Weeks-Chandler-Anderson(WCA) potential for various bond length and
densities, and compared with those in Y.-H. Shin {\it et
al.}\cite{shin01}. Our algorithm is easy to implement
and total CPU time is relatively inexpensive.
\end{abstract}


\maketitle

\section{Introduction}

One of the ways to quantify the dynamical instability of a many-body
system is to examine its Lyapunov exponents. Lyapunov exponents are
a measure of the average rate at which nearby trajectories converge
or diverge in the phase space\cite{gaspard98,hoover99,dorfman99}.
 A positive Lyapunov exponent
indicates a divergence between nearby trajectories, {\it i.e.,} a
high sensitivity to initial conditions. There are various methods to
compute Lyapunov exponents in the
literature\cite{zeng91,christiansen97,rangarajan98,ramasubramanian99,posch06}.

In this paper we propose a new method to calculate Lyapunov
exponents of the rigid diatomic molecules in three dimensions, which
does not require periodic rescaling of the bond length. Rigid
diatomic molecules are often described by hard dumbbells and there
have been several works on the Lyapunov instability for the diatomic
molecular models\cite{borzsak96,dellago97,milanovic98,milanovic98b,kum98,shin01,milanovic02}.

Our method is based on the algorithms proposed by Omelyan {\it et
al.}\cite{omelyan02}. They proposed the optimized Verlet-like
algorithms which are derived on the basis of an extended
decomposition scheme at the presence of a free parameter and which are more efficient than the original Verlet versions that
corresponds to a particular case when the introduced parameter is
equal to zero.

In the present paper, we extend Omelyan's algorithms to the
calculation of Lyapunov exponents, and show that our algorithm works
efficiently. We compare present results with those appeared in
Ref.\cite{shin01}, which uses two coordinates representations to
avoid the singularity occurring in the equations of motion by
combining with the adaptive Runge-Kutta-Fehlberg method of order
four.

\section{Equations of motion}

The position and the orientation of the diatomic molecule are
specified by a vector to the center of mass, $\vec{q}$, and a unit
vector which points along the molecular axis. We denote this unit
vector by $\vec{S}$. Then, the equations of motion are
\cite{chen01}
\beq
 \dot{\vec{q}} &=& {\vec{p} \over M}, \nonumber \\
 \dot{\vec{p}} &=& - { \partial V \over \partial \vec{q}}, \nonumber \\
  \dot{\vec{S}} &=& {1 \over I} \vec{L} \times \vec{S} =
  \vec{\omega} \times \vec{S}, \nonumber   \\
  \dot{\vec{L}} &=& - \vec{S} \times { \partial V \over \partial  \vec{S}},
\eeq
where $\vec{p}$ represents the momentum, $M$ the total mass of the
molecule, $I$ the moment of inertia, and $\vec{L}$ the angular
momentum. As in the Verlet algorithm, we split the Hamiltonian for
rigid diatomic molecular motion into kinetic and potential parts:
$ H = T + V$, where $T = {1 \over 2 M} \vec{p} \cdot \vec{p} + {1
\over 2 I} \vec{L} \cdot \vec{L}$ and $V = V (\vec{q}, \vec{S})$.
The Weeks-Chandler-Anderson(WCA) potential with a cutoff length
$r_c$ = 2$^{1/6}$ is used as the intermolecular potential
function. It should be noted that the previous work on the system
composed of hard dumbells\cite{milanovic98} employs more
complicated algorithm because of the discontinuity imposed on the
interaction potential. In the present work, all the numerical
results are based on the simulations performed on the system
composed of 32 diatomic molecules.

The exact solutions with initial condition $\vec{x}_0 = (\vec{q}_0,
\vec{p}_0, \vec{S}_0, \vec{L}_0) $ are given as follows.

1. Motion induced by $V$ :

 \beq
  {\rm exp_V(t)} \vec{x}_0 =
   \left( \vec{q}_0,
   ~\vec{p}_0 - t {\partial V \over \partial \vec{q}}
    |_{\vec{x}_0}, ~\vec{S}_0,
          ~\vec{L}_0 - t \vec{S}_0 \times { \partial V \over
          \partial \vec{S}} |_{\vec{x}_0} \right).
\eeq

2. Motion induced by $T$ :

\beq
   {\rm exp_T(t)} \vec{x}_0 = \left( \vec{q}_0 + {t \over
   M} \vec{p}_0, ~\vec{p}_0, ~\vec{S'}, ~\vec{L}_0 \right),
\eeq
 where we use the algorithm in Ref.\cite{omelyan01} to update $\vec{S}$
\beq
 \vec{S'} = { ( \vec{S}_0 + (\vec{\omega} \times \vec{S}_0)
\Delta t
        + {\Delta t^2 \over 2} [ \vec{\omega} (\vec{\omega} \cdot \vec{S}_0) - {1 \over 2}
        (\vec{\omega} \cdot \vec{\omega} ) \vec{S}_0 ] )
        \over [ 1 +  ( \omega \Delta t / 2) ^2 ] }.
\eeq
One of great advantages of this algorithm is that we don't need
to adjust the bond length during the simulations. This is also true
in Ref.\cite{shin01}, where the bond length is naturally fixed
without using any additional constraint. However, the equations of
motion in our model are much simpler than those used in Ref.
\cite{shin01}, and they are easy to implement.

A split Hamiltonian scheme for the integration, thus, is given as
\begin{widetext}
\beq
 \left( \vec{q}^{i+1}, \vec{p}^{i+1}, \vec{S}^{i+1},
\vec{L}^{i+1} \right) &=& {\rm exp_V} ({1 \over 2} \Delta t) ~{\rm
exp_T} (\Delta t)
           ~{\rm exp_V} ({1 \over 2} \Delta t)
           \times  \left(  \vec{q}^i, \vec{p}^i, \vec{S}^i, \vec{L}^i
              \right).
\eeq
\end{widetext}
In the actual simulations, however, we use the optimized Verlet-like
algorithms proposed recently by Omelyan {\it et
al.}\cite{omelyan02}to get the Lyapunov exponents. In their paper it
is shown that the optimized position-Verlet-like(OPV) algorithm is a
little more efficient than the optimized velocity-Verlet-like(OVV)
algorithm. Thus, we use the OPV algorithm in this paper. The
algorithm reads

\beq
 \vec{v}_I &=& \vec{v}(t) + {1 \over m} \vec{f}(\vec{r}(t)) \xi \Delta t,
 \nonumber \\
 \vec{r}_I &=& \vec{r}(t) + \vec{v}_I \Delta t/2, \nonumber \\
\vec{v}_{II} &=& \vec{v}_I(t) + {1 \over m} \vec{f}(\vec{r}_I)
(1-2\xi) \Delta t,
 \nonumber \\
\vec{r} (t+\Delta t) &=& \vec{r}_I(t) + \vec{v}_{II} \Delta t/2, \nonumber \\
 \vec{v} (t+\Delta t) &=& \vec{v}_{II} + {1 \over m} \vec{f}(\vec{r}(t+\Delta t)) \xi \Delta
 t,
\eeq
 where the parameter $\xi \approx$ 0.1931833275037836 and
 $\vec{r}_I, \vec{r}_{II}, \vec{v}_I$, and $\vec{v}_{II}$ denote
the center of positions and their velocities of a diatomic
molecule. Similarly, we have the propagation of the unit vector
$\vec{S}$ and angular momentum $\vec{L}$ from time $t$ to
$t+\Delta t$.
A time step $\Delta t$ = 0.0005 is used in our simulations.

\section{Lyapunov exponents}

The initial configuration of the molecules and the simulation method
in Ref.\cite{shin01} are used in our simulations. The velocities are
repeatedly scaled to adjust the required temperature from
sufficiently high temperature . Once the required temperature is
obtained, the system is equilibrated for 500 time units (10$^6$
iterations), and data is collected for 500 time units(10$^6$
iterations) to evaluate the Lyapunov exponents. Below all
quantities are given in reduced units.

In order to get the exponents we use the classical method of
Benettin {\it at al.}\cite{benettin76} refined by Hoover and
Posch\cite{hoover85,hoover87,posch88,posch89} that requires
continuous orthonormalization. The corresponding equations for the
variation are given as follows.
\beq
 \delta \dot{\vec{q}} &=& {\delta \vec{p} \over M}, \nonumber \\
 \delta \dot{\vec{p}} &=& - \delta { \partial V \over \partial \vec{q}}, \nonumber \\
 \delta \dot{\vec{S}} &=& {1 \over I} \delta (\vec{L} \times \vec{S}) =
  \delta(\vec{\omega} \times \vec{S}), \nonumber   \\
 \delta \dot{\vec{L}} &=& - \delta(\vec{S} \times { \partial V \over \partial
 \vec{S}}).
\eeq
It should be noted that we consider only the first order,
$\cal{O}$($\Delta t$), for updating $\delta \vec{S}$. In our model
we have $12N$ dimensional phase space. In order to compare our
results with those from $10N$ dimensional phase space in
Ref.\cite{shin01}, we subtract the additional two degrees of
freedoms in each diatomic molecule. We have two constraints in
equations of motion : \beq
  \vec{S} \cdot \delta \vec{S} &=& 0   ,
\label{eq_const1}
 \eeq
 \beq
  \vec{S} \cdot \delta \vec{L} + \delta \vec{S} \cdot \vec{L} &=& 0.
 \label{eq_const2}
  \eeq
These equations are obtained from two conditions : $\vec{S} \cdot \vec{S}$ = 1
(normalization) and $\vec{S} \cdot \vec{L}$ = 0 (orthogonality).
In order to fulfill two constraints $\delta \vec{S}$ and $\delta \vec{L}$ are replaced with,
\beq
 \delta \vec{S} \equiv \delta \vec{S} - (\vec{S} \cdot \delta \vec{S})
 \vec{S} ,
\label{eq_newS}
\eeq
\beq
   \delta \vec{L} \equiv \delta \vec{L} - (\vec{S} \cdot \delta \vec{L} + \delta \vec{S} \cdot \vec{L}) \vec{S} ,
\label{eq_newL}
\eeq
respectively. Then, $12N$ dimensional phase space can be reduced to
$10N$ dimensional one.

\begin{figure}
\includegraphics[width=3.0in]{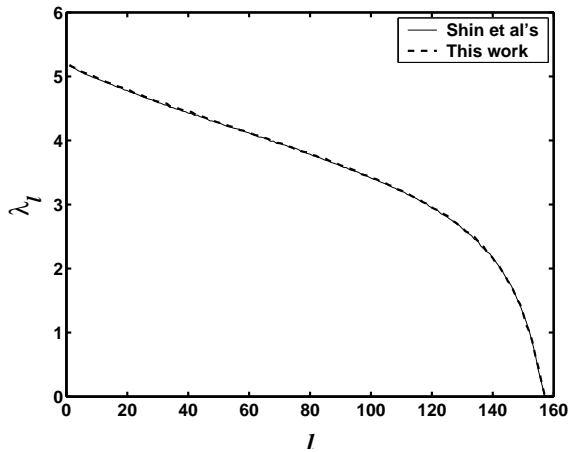}
\caption{Comparison of the Lyapunov exponents at $T$=1.0, $D$=0.3, $B$=1.0.}
\label{fig1}
\end{figure}

We find the characteristic features of the spectra of
the Lyapunov exponents in our model are the same as those in
Ref.\cite{shin01}. In Fig. \ref{fig1} we compare the positive branch
of exponents from our calculation with that from Ref.\cite{shin01}
at $T$=1.0, $D$=0.3, $B$=1.0, where $T$ is temperature, $D$ is the
density, and $B$ is the bond length, respectively. $\lambda_l$ is
the discrete spectrum of the Lyapunov exponents, and index $l$
represents 1, ..., 160, {\it i.e.} the half of total number of all
phase space variables .

\begin{figure}
\includegraphics[width=3.0in]{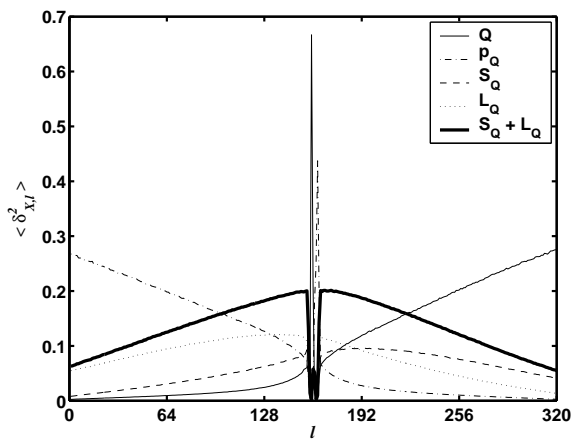}
\caption{Mean-squared values of the projection at $T$=1.0, $D$=0.3, $B$=1.0.}
\label{fig2}
\end{figure}

\begin{figure}
\includegraphics[width=3.0in]{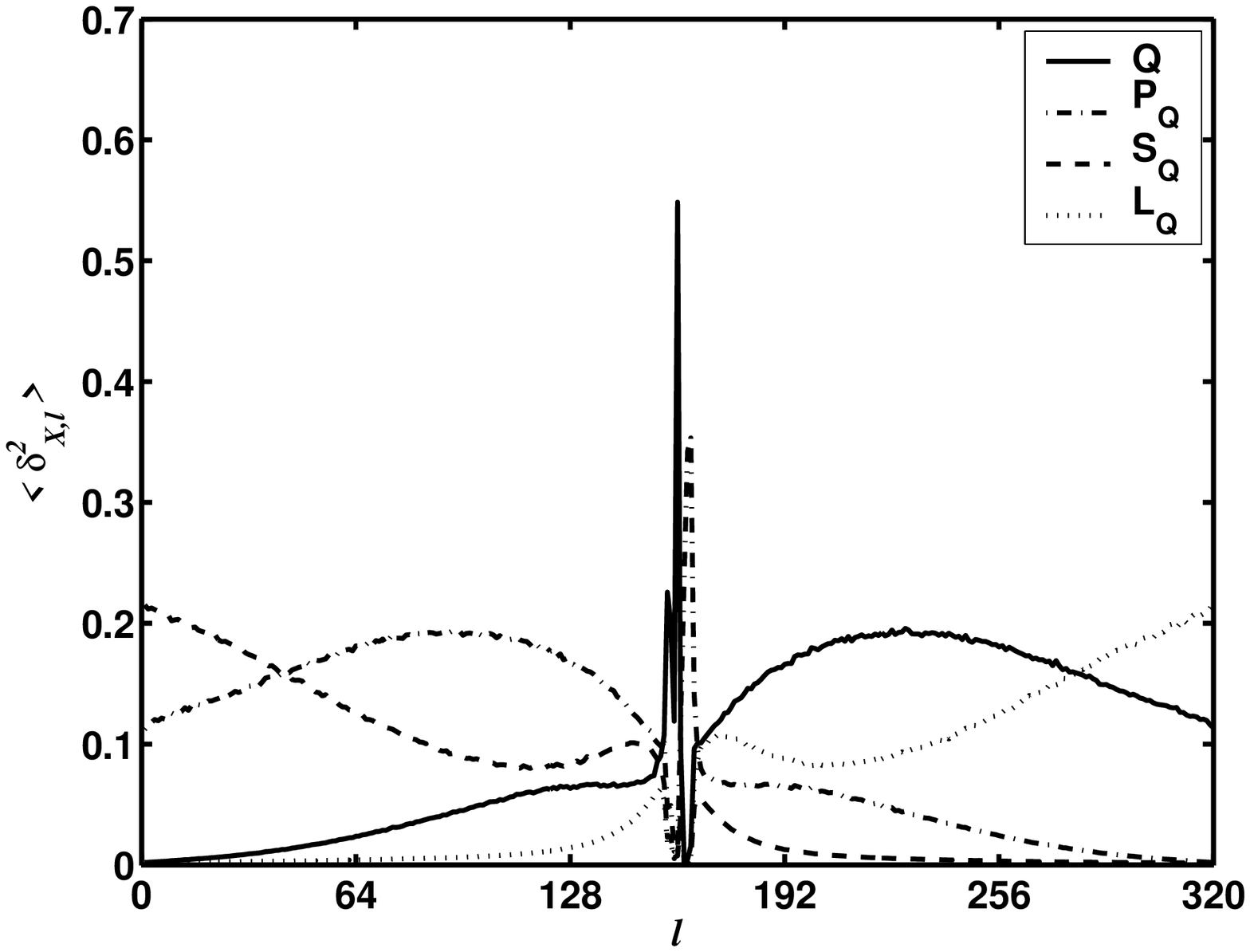}
\includegraphics[width=3.0in]{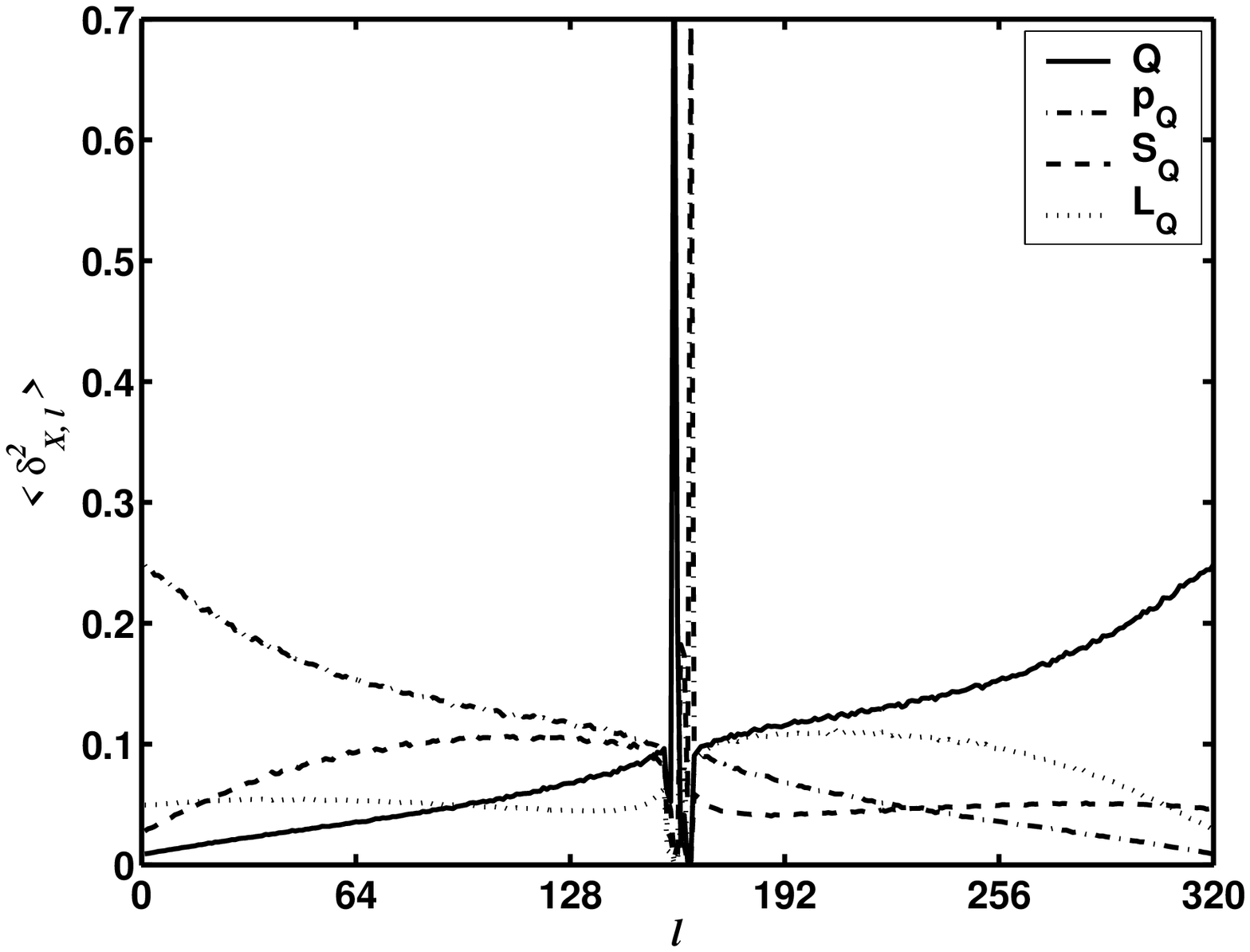}
\caption{Same as in Fig.\ref{fig2}, but at $T$=1.0, $D$=0.3, $B$=0.2
(top), and at $T$=0.1, $D$=0.3, $B$=1.0 (bottom).} \label{fig3}
\end{figure}

\begin{table}
\caption{The largest Lyapunov exponent $\lambda_1$ and the smallest positive Lyapunov exponent $\lambda_{156}$,
and four vanishing exponents ($\lambda_{157}$, $\lambda_{158}$, $\lambda_{159}$, and $\lambda_{160}$)
for various bond length B at a fixed number density D=0.3 (upper) and D=0.5 (lower).}
\begin{tabular}{c|ccccccccc}
\hline \hline
 B              &  0.2    &&  0.4   && 0.6   &&  0.8   && 1.0                   \\ \hline
$\lambda_1$      &  4.320 &&  4.574 && 4.791 &&  4.989 && 5.179                 \\
$\lambda_{156}$  &  0.104 &&  0.170 && 0.274 &&  0.272 && 0.229                 \\
$\lambda_{157}$  &  0.025 &&  0.046 && 0.041 &&  0.034 && 0.030                 \\
$\lambda_{158}$  &  0.018 &&  0.014 && 0.014 &&  0.017 && 0.013                 \\
$\lambda_{159}$  &  0.010 &&  0.008 && 0.012 &&  0.012 && 0.011                 \\
$\lambda_{160}$  &  0.012 &&  0.009 && 0.009 &&  0.011 && 0.007                 \\
\hline \hline
\end{tabular}

\vspace{0.2cm}
\begin{tabular}{c|ccccccccc}
\hline \hline
 B               &  0.2    &&  0.4   && 0.6   &&  0.8   && 1.0                   \\ \hline
$\lambda_1$      &  6.158  &&  6.526 && 6.557 &&  6.327 && 5.746                 \\
$\lambda_{156}$  &  0.108  &&  0.113 && 0.052 &&  0.052 && 0.032                 \\
$\lambda_{157}$  &  0.068  &&  0.056 && 0.028 &&  0.015 && 0.013                 \\
$\lambda_{158}$  &  0.020  &&  0.013 && 0.015 &&  0.014 && 0.011                 \\
$\lambda_{159}$  &  0.014  &&  0.011 && 0.010 &&  0.011 && 0.012                 \\
$\lambda_{160}$  &  0.010  &&  0.009 && 0.007 &&  0.010 && 0.003                 \\
\hline \hline
\end{tabular}
\label{table1}
\end{table}

Table \ref{table1} shows the largest Lyapunov exponents $\lambda_1$,
the smallest positive Lyapunov exponent $\lambda_{156}$, and four
vanishing exponents ($\lambda_{157}$, $\lambda_{158}$,
$\lambda_{159}$, and $\lambda_{160}$) for various bond length B at a
fixed number density D=0.3, and D=0.5, respectively. Temperature is
set to $T$=1.0 for both cases. The same trend for the largest
Lyapunov exponent $\lambda_1$ was already shown in
Ref.\cite{shin01}.

On the other hand, the dynamics of the tangent vectors in the
subspace are slightly different. We calculate the mean-squared value
of the projection of tangent vectors $\delta_l$ onto $TX$ subspace
which is defined as usual
\beq
  \la \vec{\delta}^2_{X,l} \ra = \la \mathcal{P}(X) \vec{\delta}_l
  \cdot \mathcal{P}(X) \vec{\delta}_l \ra .
\eeq
Fig.\ref{fig2} shows the projection at $T$=1.0, $D$=0.3, $B$=1.0,
where $Q$ denotes $x$, $y$, and $z$ components.
In our model $\la \vec{\delta}^2_{X,l} \ra$ for $\vec{S}$ and
$\vec{L}$ phase space and their tangent space $\delta \vec{S}$ and
$\delta \vec{L}$ are not symmetric with respect to the center.
A similar asymmetric behavior was also found in Ref.\cite{hoover01}.
However, we find the Hamiltonian nature of the system, {\it i.e.},
an increase of instability accumulated in one subspace is always
accompanied with a decrease of instability in its conjugate
subspace. In Fig.\ref{fig2} we also show $S_Q + L_Q$, which is symmetric with respect to the center,
for comparison.

\begin{figure}[t]
\includegraphics[width=3.0in]{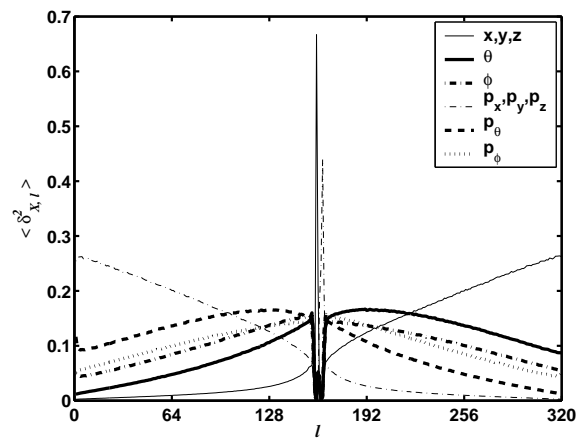}
\caption{Same as in Fig. \ref{fig2}. After transforming of the
coordinate system.} \label{fig4}
\end{figure}

It is interesting to note that the overall patterns are symmetric if
temperature is low or the bond length is small. In Fig.\ref{fig3} we
show the projection at $T$=1.0, $B$=0.2, and at $T$=0.1, $B$=1.0,
respectively. The density is set to $D$=0.3 for both cases. It is
not clear to us why the projection depends on temperature or the bond
length, and symmetry becomes broken as temperature or the bond
length is increasing.
We speculate that this might come from a
relation between $\vec{S}$ and $\vec{L}$. Note that the correct
conjugate momentum for rotation is\cite{chen01}
\beq
\vec{\Pi} = I \dot{\vec{S}} = \vec{L} \times \vec{S},
\eeq
not simply $\vec{L}$. More detailed analyses will be needed
regarding the asymmetric behavior. It can be shown that the overall
patterns are always symmetric if our coordinate system is
transformed to that of Ref.\cite{shin01}, {\it i.e.}, ($\vec{q}$,
$\vec{p}$, $\theta$, $\phi$, $p_{\theta}$, $p_{\phi}$) (See Fig
\ref{fig4}). Clearly, this shows that our coordinate system is
consistent with that of Ref.\cite{shin01}.

In Table \ref{table2} we show comparison of total CPU time for our
simulations and Shin {\it et al.}'s on the Lyapunov exponents. Our code is relatively
inexpensive, although we need to implement two constraints in
Eq.(\ref{eq_const1}) and Eq.(\ref{eq_const2}). In Ref.\cite{shin01}
the coordinate transformation is needed to avoid singularity
occurring in the equations of motion, and a sophisticated integrator
is used because the coordinate transformation cannot be applied to the
calculation of the Lyapunov exponents. These might consume a
relatively large CPU time during the simulations.

\begin{table}[t!]
\caption{Comparison of total CPU time(in hours) for the simulations
on a single node (Intel 3.06GHz CPU).}
\begin{tabular}{c|ccccccccc}
\hline \hline
                        &  This work(A)   &&  Shin {\it et al}'s(B)    &&    A/B   \\ \hline
T=1.0, D=0.3, B=0.2     &    8.8h     &&      16.5h            &&     0.53    \\
T=1.0, D=0.3, B=1.0     &    15.5h     &&      24.6h            &&    0.63     \\
T=1.0, D=0.5, B=0.2     &    17.1h      &&      28.3h           &&    0.60     \\
\hline \hline
\end{tabular}
\label{table2}
\end{table}


In summary, we propose a new approach to calculate Lyapunov exponents of the rigid diatomic molecules
in three dimensions, which does not need a rescaling of the bond length,
and it is computationally relatively inexpensive.

The authors would like to thank Dr. Dong-Chul Ihm for contributing to
the early stages of this work. S.C. is grateful to Prof. H.A. Posch for
helpful comments in the early stage of this work, and he also thanks
Dr. C.-H. Cho and Dr. Y.-H. Shin for useful discussions.
This work was supported by the Korea Research Foundation Grant No. KRF-2005-070-C00065.



\begin{thebibliography}{99}

\bibitem{shin01}
Y.-H. Shin, D.-C. Ihm, and E.-K. Lee,
 Phys. Rev. E \textbf{64}, 041106 (2001).

\bibitem{gaspard98}
P. Gaspard,
 {\it Chaos, Scattering and Statistical Mechanics}, Cambridge University Press,
 Cambridge, 1998.

 \bibitem{hoover99}
 W.G. Hoover,
 {\it Time Reversibility, Computer Simulation, and Chaos},
 World Scientific, Singapore, 1999.

 \bibitem{dorfman99}
 J.R. Dorfman,
 {\it An Introduction to Chaos in Nonequilibrium Statistical Mechanics},
 Cambridge University Press, Cambridge, 1999.


\bibitem{zeng91}
X. Zeng, R. Eykholt, and R.A. Pielke,
 Phys. Rev. Lett. \textbf{66}, 3229 (1991).

\bibitem{christiansen97}
 F. Christiansen and H.H. Rugh,
 Nonlinearity \textbf{10}, 1063 (1997).

\bibitem{rangarajan98}
 G. Rangarajan, S. Habib, and R.D. Ryne,
 Phys. Rev. Lett. \textbf{80}, 3747 (1998).

\bibitem{ramasubramanian99}
 K. Ramasubramanian and M.S. Sriram,
  Phys. Rev E \textbf{60}, R1126 (1999).

 \bibitem{posch06}
 H.A. Posch and W.G. Hoover,
 J. Phys. : Conference Series \textbf{31}, 9 (2006); and references therein.

\bibitem{borzsak96}
 I. Borzs\'{a}k, H.A. Posch, and A. Baranyai,
 Phys. Rev. E \textbf{53}, 3694 (1996).

 \bibitem{dellago97}
 Ch. Dellago and H.A. Posch,
 Physica A \textbf{237}, 95 (1997).


 \bibitem{milanovic98}
 L. Milanovi\'{c}, H.A. Posch, and W.G. Hoover,
 Chaos \textbf{8}, 455 (1998).

 \bibitem{milanovic98b}
 L. Milanovi\'{c}, H.A. Posch, and W.G. Hoover,
 Molec. Phys. \textbf{95}, 281 (1998).



 \bibitem{kum98}
O. Kum, Y.H. Shin, E.K. Lee,
 Phys. Rev. E \textbf{58}, 7243 (1998).


 \bibitem{milanovic02}
 L. Milanovi\'{c}, H.A. Posch,
 J. Molecular Liquids \textbf{96-97}, 221 (2002).


\bibitem{omelyan02}
 I.P. Omelyan, I.M. Mryglod, and R. Folk,
 Phys. Rev. E \textbf{65}, 056706 (2002).

\bibitem{chen01}
 B. Chen and J.H. Schenker,
  Applied Numerical Mathematics \textbf{38}, 21 (2001).

\bibitem{omelyan01}
 I.P. Omelyan, I.M. Mryglod, and R. Folk,
 Phys. Rev. Lett. \textbf{86}, 898 (2001).

\bibitem{benettin76}
G. Benettin, L. Galgani, and J.-M. Strelcyn,
  Phys. Rev. A \textbf{14}, 2338 (1976).

\bibitem{hoover85}
W.G. Hoover and H.A. Posch,
 Phys. Lett. A \textbf{113}, 82 (1985).

  \bibitem{hoover87}
W.G. Hoover and H.A. Posch,
 Phys. Lett. A \textbf{123}, 227 (1987).


\bibitem{posch88}
H.A. Posch and W.G. Hoover,
 Phys. Rev. A \textbf{38}, 473 (1988).

\bibitem{posch89}
H.A. Posch and W.G. Hoover,
Phys. Rev. A \textbf{39}, 2175 (1989).

\bibitem{hoover01}
W.G. Hoover, H.A. Posch, and C.G. Hoover,
 J. Chem. Phys. \textbf{115}, 5744 (2001).

\end{thebibliography}
\end{document}